\documentclass[12pt]{article}

\usepackage{epsf}
\usepackage{color}
\def\note#1{{\color{red} #1}}

\usepackage{amsmath,amssymb,color,graphicx,graphics,amscd,amsfonts}

\begin{document}

\begin{flushright} 
 OIQP-16-04
\end{flushright} 

\begin{Large}
\vspace{1cm}
\begin{center}
{\bf Noncommutative Frobenius algebras and open-closed duality} \\ 
\end{center}
\end{Large}

\vspace{1cm}

\begin{center}
{\large Yusuke Kimura}

\vspace{0.3cm}

Okayama Institute for Quantum Physics (OIQP), \\ 
Furugyocho 1-7-36, Naka-ku, Okayama, 703-8278, Japan\\ {\tt londonmileend\_at\_gmail.com}
   \\ 


\end{center}

\vspace{1cm}

\begin{abstract}
\noindent

Some equivalence classes in symmetric group lead to 
an interesting class of noncommutive 
and associative algebras. 
From these algebras we construct noncommutative Frobenius algebras.
Based on the correspondence between Frobenius algebras and 
two-dimensional topological field theories, 
the noncommutative Frobenius algebras can be interpreted as 
topological open string theories. It is observed that 
the centre of the algebras are related to 
closed string theories via open-closed duality.
Area-dependent 
two-dimensional field theories are also studied.

\end{abstract}


\section{Introduction}

Topological field theories are the simplest field theories 
whose  
amplitudes are indepedendent of the local Riemann structure.
Two-dimensional topological field theories may 
attract a particular interest because they 
can be regarded as toy models of string 
theory.  
If we ignore the conformal structure of Riemann surfaces corresponding to 
worldsheets, 
we have two-dimensional topological field theories 
\cite{Segal_CFT,Atiyah}.

In this paper we will focus on the one-to-one correspondence 
between  
two-dimensional topological field theories and 
Frobenius algebras \cite{9212154,9205031,Abrams,J.Kock}. 
This may suggest an interesting story that 
two-dimensional topological field theories can be re-constructed from a certain kind of 
algebras, which reminds us of some recent approaches of string theory that   
it is re-constructed from discrete data like matrices in 
AdS/CFT correspondence and the matrix models.

A Frobenius algebra is defined as a finite-dimensional associative algebra 
equipped with a non-degenerate bilinear form. 
Given an associative algebra, it can be a Frobenius algebra by specifying the 
bilinear form. 
It is emphasised that 
being Frobenius is not a property of the algebra, but it is a structure we provide.
We can construct different Frobenius algebras 
by giving different Frobenius structures, even though the underlying algebra 
is the same.  
For a Frobenius algebra, we can associate 
the bilinear form and the structure constant with the 
propagator and the three-point vertex of a two-dimensional 
topological field theory. 
Topological properties of the theory can be explained from 
the fact that it is an associative algebra with a non-degenerate 
bilinear form.

\quad

Our interest on the equivalence comes from the recent development 
that correlation functions of the Gaussian matrix models, encoding 
the space-time independent part of correlation functions of 
${\cal N}=4$ Super Yang-Mills, 
are expressed in terms of elements in associative algebras 
obtained from symmetric group.

In section \ref{sec:equivalence_classes} we introduce some associative algebras 
obtained from the group algebra of the symmetric group.
Sums of all elements in the conjugacy classes 
are central elements in the group algebra, and the product is closed. 
They form an associative 
and commutative 
algebra.  
We can also construct more general class of algebras 
by introducing 
equivalence relations in terms of elements 
in subgroups of the symmetric group \cite{0711.0176,0709.2158,0801.2061}. 
Such algebras are associative but in general noncommutative. 
Because the noncommutativity 
is related to  
the restriction from the symmetric group to the subalgebra, 
the noncommutative structure is sometimes referred to as restricted structure. 
Some algebras with restricted structure 
were recently studied in \cite{1601.06086}.

In section \ref{sec:review_matrix_model}, we will explain how 
the algebras introduced in section \ref{sec:equivalence_classes} 
appear 
in the description of 
gauge invariant operators in ${\cal N}=4$ Super Yang-Mills.  
The elements in the algebras
have an one-to-one correspondence with   
the gauge invariant operators.
Two-point functions of the gauge invariant operators
give a map from two elements in the algebra to a $c$-number.
Regarding the map as a bilinear form, 
we can introduce Frobenius algebras naturally from the matrix models. 
The equivalence between Frobenius algebras and 
two-dimensional topological field theories leads to an interpretation of  
correlation functions of the matrix models
in terms of two-dimensional topological field theories \cite{1301.1980}. 

In section \ref{sec:simplest_open_closed_TFT}, 
we will review concretely how Frobenius algebras describe 
topological field theories. 
Commutative Frobenius algebras correspond to topological closed string theories, 
while 
noncommutative Frobenius algebras can be associated 
with open string theories \cite{9212154,9205031,Segal_01,0609042}. 
Because the closed propagator is related to a non-planar 
one-loop diagram of an open string theory, 
the closed string theory and the open string theory should be 
described in a unified algebraic framework.
Our review will be given from this point of view. 

Section \ref{sec:restricted_algebra} will be devoted to a detailed study of 
algebras with the restricted structure from the point of view of 
the open-closed duality. 
In our previous paper \cite{1403.6572}
correlation functions of topological field theories related to 
noncommutative Frobenius algebras were studied. 
In this paper, we reinterpret the noncommutative Frobenius algebras 
as open string theories. 
Elements in the noncommutative algebras are related to open string states, while 
the centre of the algebra describes closed string states. 
The restricted structure may cause 
an area-dependence in correlation functions of the 
closed string theories. 

In section \ref{summary} we summarise this paper. 
In appendices, we give formulas, detailed computations and additional explanations. 
Appendices of our previous paper \cite{1403.6572} may also be helpful to 
check equations of this paper.

\qquad

This paper is based on the talk 
`` Noncommutative Frobenius algebras and open strigns,''
given in 
workshop `` Permutations and Gauge String duality,'' held at  
Queen Mary, University of London at July 2014. 
The relationship to our recent work \cite{1608.03188} is also briefly mentioned. 


\section{Equivalence classes in symmetric group}
\label{sec:equivalence_classes}

We consider some types of equivalence classes in 
symmetric group $S_n$. 
One is the conjugacy classes. 
The others are introduced by 
considering subgroups, which leads to an interesting class of algebras.

\quad 

Consider the group algebra of $S_n$ over $\mathbb{C}$, which is called 
$\mathbb{C}[S_n]$.  
For elements in $S_n$, 
introduce 
the equivalence relation $\sigma \sim g \sigma g^{-1}$ by any element $g$ in $S_n$. 
Because of the equivalence relation,   
the set of elements in $S_n$ are classified into conjugacy classes. 
It is then convenient to consider 
\begin{eqnarray}
[\sigma]
=\frac{1}{n!}\sum_{\tau \in S_n}\tau \sigma \tau^{-1}.
\label{sum_conjugacy_class}
\end{eqnarray}
Elements in a conjugacy class give the same value of $[\sigma]$, i.e. we have 
\begin{eqnarray}
[\sigma]=[g\sigma g^{-1}]
\end{eqnarray}
for any $g$ in $S_n$. 
In fact $[\sigma]$ is   
the sum of all elements in the conjugacy class the $\sigma$ belongs to.  
The sums are central elements in $\mathbb{C}[S_n]$, 
\begin{align}
[\sigma]\tau=\tau [\sigma], \quad 
[\sigma][\tau]=[\tau][\sigma]
\end{align}
for any $\sigma,\tau\in S_n$. 
Because $\sigma$ and $\sigma^{-1}$ are in the same conjugacy class, 
we also have 
\begin{align}
[\sigma]=[\sigma^{-1}].
\end{align}
The product of two central elements gives a central element of the form (\ref{sum_conjugacy_class}), 
\begin{align}
[\sigma_1][\sigma_2]=\frac{1}{n!}\sum_{\tau \in S_n}\tau \sigma_1 [\sigma_2] \tau^{-1}.
\end{align}
As above 
the $[\sigma]$ span a basis of the centre of the group algebra. 
Central elements form a closed commutative and associative algebra,   
which is denoted by $Z(\mathbb{C}[S_n])$.

\quad

We next introduce the equivalence relation determined by  
a subgroup $H$ of $S_n$ as 
$\sigma \sim h\sigma h^{-1}$, where 
$h\in H$ and $\sigma\in S_{n}$. 
Generalising (\ref{sum_conjugacy_class}), 
the equivalence classes under the subgroup $H$ are classified by 
\begin{align}
[\sigma]_{H}=\frac{1}{|H|}\sum_{h\in H} h\sigma h^{-1},
\label{elements_restriction_algebra}
\end{align}
where $|H|$ denotes the dimension of the subgroup. 
They have the following properties  
\begin{align}
[\sigma]_H=[h \sigma h^{-1}]_H, \quad 
h[\sigma]_H=[\sigma]_H h \quad (h\in H),
\label{commuting_algebra_restriction_H}
\end{align}
and also  
\begin{align}
[\sigma]_H\neq [\sigma^{-1}]_H. 
\label{restricted_sum_inverse}
\end{align}
The product of 
two elements 
of the form (\ref{elements_restriction_algebra}) can be expressed
by the form (\ref{elements_restriction_algebra}), 
\begin{align}
[\sigma_1]_{H}[\sigma_2]_H=\frac{1}{|H|}\sum_{h\in H} h\sigma_1[\sigma_2]_H h^{-1}.
\end{align}
The closed algebra formed by 
the elements (\ref{elements_restriction_algebra}) is called 
$\mathbb{C}[S_{n}]_H$ in this paper. 
Our notation is that 
$\mathbb{C}[S_{n}]_H=Z(\mathbb{C}[S_{n}])$
for $H=S_{n}$.
A big difference from the algebra relevant for the conjugacy classes is that 
the elements given in 
(\ref{elements_restriction_algebra}) span a noncommutative algebra, 
\begin{eqnarray}
[\sigma]_H[\tau]_H \neq [\tau]_H[\sigma]_H.
\label{noncommutative_restriction}
\end{eqnarray}

There are several interesting subgroups we can choose for $H$. 
For instance we choose
$H=S_{n_1}\times S_{n_2}\times \cdots$, 
where $n_1+n_2=\cdots =n$ \cite{0711.0176,0801.2061,0807.3696}. 
It is also interesting to consider the wreath product group $H=S_{n/2}[S_2]$, 
which has played a role 
recently in \cite{1608.03188}.

\quad 

Brauer algebras can also be similarly considered. 
The walled Brauer algebra $B(n_1,n_2)$ with the subgroup 
$H=S_{n_1}\times S_{n_2}$ 
gives equivalence classes determined by 
\begin{eqnarray}
[b]_{H}=
\frac{1}{n_1!n_2!}\sum_{h\in S_{n_1}\times S_{n_2}}h bh^{-1},
\end{eqnarray}
where $b$ is an element in the walled Brauer algebra. 
These quantities can classify 
a certain class of gauge invariant operators \cite{0709.2158,1206.4844}.

\quad 

Before closing this section, we shall give a remark about another basis 
labelled by a set of representation labels. 
We may use the projection operators 
as a basis of $Z(\mathbb{C}[S_n])$. 
The change of basis is given by 
\begin{eqnarray}
[\sigma]=\sum_{R\vdash n} \frac{1}{d_R}\chi_R(\sigma)P^R, 
\label{central_element_expand_projector}
\end{eqnarray}
where $P^R$ and $\chi_R$ are 
the projection operator and the character 
associated with an irreducible representation $R$, 
and 
$d_R$ is the dimension of $R$ of $S_n$. 
The sum is over all Young diagrams with $n$ boxes. 
The projectors by definition satisfy 
\begin{eqnarray}
P^R P^S=\delta_{RS}P^R.
\end{eqnarray}
In this paper we call this basis {\bf a representation basis}.

\quad 

The algebras with the restricted structure also 
admit a similar expansion, 
\begin{eqnarray}
[\sigma]_H=\sum_{R,A,\mu,\nu}\frac{1}{d_A}
\chi^R_{A,\mu\nu}([\sigma]_H)P^R_{A,\mu\nu}, 
\label{expansion_restricted_operator}
\end{eqnarray}
where $R$ is an irreducible representation of $S_n$ and 
$A$ is an irreducible representation of $H$. 
The $P^R_{A,\mu\nu}$ are obtained by decomposing the projector $P^R$ 
\begin{eqnarray}
P^R=\sum_{A}P^R_A=\sum_{A,\mu}P^R_{A,\mu\mu}.
\end{eqnarray}
The indices $\mu,\nu$ behave like matrx indices, 
\begin{align}
P^R_{A,\mu\nu}P^{R^{\prime}}_{A^{\prime},\rho\sigma}
=\delta_{RR^{\prime}}\delta_{AA^{\prime}}\delta_{\nu\rho}
P^{R}_{A,\mu\sigma},
\label{projector_like_matrix}
\end{align}
and the trace part 
with respect of these indices is denoted by 
$P^R_A$. 
Let $M^R_A$ be the number of times $A$ appears in $R$ 
under the embedding of $H$ in $S_n$. 
We call $\mu,\nu$  multiplicity indices 
because they run over $1,\cdots,M^R_A$. 
See \cite{0411205,0701066,0801.2061,0807.3696} for the construction of the operators $P^R_{A,\mu\nu}$ 
and the coefficients $\chi^R_{A,\mu\nu}$
for $H=S_{n_1}\times \cdots  \times S_{n_2}$. 
The noncommutativity in (\ref{noncommutative_restriction}) 
is
reflected to the noncommutativity of 
the matrix structure in (\ref{projector_like_matrix}). 
The $P^R_{A,\mu\nu}$ commute 
with all elements in the subgroup, 
\begin{eqnarray}
P^R_{A,\mu\nu} h=hP^R_{A,\mu\nu}, \quad h\in H.
\end{eqnarray}
In general there may exist some representation bases 
in the algebras.
Elements in $\mathbb{C}[S_n]_H$ allow a different expansion from 
(\ref{expansion_restricted_operator}), which will be given in 
Appendix \ref{TFT_covariant_basis}.


\section{Frobenius algebras and matrix models}
\label{sec:review_matrix_model}

In the previous section
we have introduced some equivalence relations in the symmetric group. 
The equivalence classes also play a role 
in the description of gauge invariant operators 
in ${\cal N}=4$ SYM, where it can be found that 
the number of the equivalence classes is equal to the number of 
the gauge invariant operators. 
Two-point functions of the gauge theory 
naturally give a map from $W\otimes W \rightarrow \mathbb{C}$, 
where $W$ is a basis of the algebra, 
and the map can be adopted to make the algebra Frobenius. 

\quad 

Consider the matrix model
\begin{eqnarray}
\langle f(X,X^{\dagger})\rangle =
\int \prod_a [dX_a dX_a^{\dagger}]
e^{-2tr(X_a X_a^{\dagger})}
f(X,X^{\dagger}), 
\label{matrix_integral}
\end{eqnarray}
where the measure is normalised to give 
\begin{eqnarray}
\langle (X_a)_{ij}(X_b^{\dagger})_{kl} \rangle 
=\delta_{il}\delta_{kl}\delta_{ab}. 
\end{eqnarray}
Here $f$ is an invariant quantity under 
$X_a \rightarrow gX_a g^{-1}$, 
$X_a^{\dagger} \rightarrow g X_a g^{-1}$ 
where $g\in U(N)$, and $N$ is the size of the matrices. 
Such invariant quantities are in general 
linear combinations of multi-traces. 
They are called gauge invariant operators. 

When we say holomorphic operators, 
we consider (linear combinations of) multi-traces built from only $X_a$. 
By two-point functions of holomorphic operators, 
we mean 
\begin{eqnarray}
\langle f(X)g(X^{\dagger})\rangle =
\int \prod_a [dX_a dX_a^{\dagger}]
e^{-2tr(X_a X_a^{\dagger})}
f(X)g(X^{\dagger}). 
\label{2pt_matrix_integral}
\end{eqnarray}
As a shorthand, we denote it by $\langle fg \rangle$.

\quad 

We first consider holomorphic gauge invariant operators built from 
one kind of complex matrix. 
We use $n$ to 
express  
the number of matrices involved in the operators. 
For example at $n=3$ we have three operators, 
$tr(X^3)$, $tr(X^2)trX$, $(trX)^3$. 
Counting the number of gauge invariant operators is 
a combinatorics problem, and 
it has been known that the number of matrix gauge invariant operators 
involving $n$ $X$'s
is equivalent to the number of conjugacy classes 
of the symmetric group $S_n$.\footnote{
At $n=3$ 
we have three multi-traces 
$tr(X^3)$, $trXtr(X^2)$ and $(trX)^3$, 
while we have three conjugacy classes in $S_3$, $[1,1,1]$, $[1,2]$ and $[3]$. 
Each conjugacy class corresponds to an integer partition of $n$. 
This correspondence does not work when 
the matrix size $N$ is less than $n$. 
For $N=2$, 
we have the identity among the operators, $tr(X^3)-\frac{3}{2}trXtr(X^2)+\frac{1}{2}(trX)^3=0$, 
which means that 
only two of the three operators 
are independent. This is an example of finite $N$ constraints. 
Operators with finite $N$ constraints taken into account for $n>N$ 
can be correctly counted by considering a representation basis. 
\label{footnote_finite_N_constraint}
}

From the correspondence, it is expected that elements in $Z(\mathbb{C}[S_n])$ can 
classify the gauge invariant operators.
Consider  
\begin{eqnarray}
O(\sigma):=(X)_{i_1}^{i_{\sigma(1)}}(X)_{i_2}^{i_{\sigma(2)}}
\cdots (X)_{i_n}^{i_{\sigma(n)}}.
\end{eqnarray}
Regarding $X$ as an endormorphism acting on a vector space $V$, we can express 
it as 
\begin{eqnarray}
O(\sigma)=tr_n(\sigma X^{\otimes n}),
\end{eqnarray}
where the $\sigma$ permutes the $n$ factors in $V^{\otimes n}$ and 
$tr_n$ is a trace over the tensor space $V^{\otimes n}$.  
We can show that 
\begin{eqnarray}
O(\sigma)=O(\tau\sigma \tau^{-1})
\end{eqnarray}
for any $\tau \in S_{n}$, then we have 
\begin{eqnarray}
O([\sigma])=\frac{1}{n!}\sum_{\tau\in S_n}O(\tau\sigma \tau^{-1})=O(\sigma).
\end{eqnarray}

Two-point functions of 
the gauge invariant operators are computed as
\begin{eqnarray}
\langle O([\sigma]) O([\tau]) \rangle =tr^{(r)}(\Omega_n [\sigma][\tau]).
\label{twopt_single_matrix}
\end{eqnarray}
Here the trace is a trace over regular representation, 
\begin{eqnarray}
tr^{(r)}(\sigma)=\sum_{R\vdash n}d_R \chi_R(\sigma), 
\label{2pt_RHS_trace}
\end{eqnarray}
where $d_R$ is the dimension of an irreducible representation $R$ 
of symmetric group $S_n$, and 
$\chi_R$ is the character associated with $R$. The sum is over all Young diagrams 
with $n$ boxes. 
This derivation is reviewed in section 2 of our previous paper \cite{1403.6572}.
The $N$-dependence is entirely contained in the Omega factor 
\begin{align}
\Omega_{n}
=\sum_{\sigma\in S_n}\sigma N^{C_{\sigma}-n}
 =1+\frac{1}{N^n}\sum_{\sigma\neq 1}\sigma N^{C_{\sigma}},
\label{Omega_factor}
\end{align}
where $C_{\sigma}$ counts the number of cycles in the $\sigma$.

The right-hand side of (\ref{twopt_single_matrix}) can also be written as 
\begin{eqnarray}
tr^{(r)}(\Omega_n [\sigma][\tau])
=\frac{n!}{N^n}\sum_R \frac{DimR}{d_R}\chi_R(\sigma) \chi_R(\tau). 
\label{bilinear_map_two_pt_single_matrix}
\end{eqnarray}
Here 
$DimR$ is the dimension of an irreducible representation $R$ of $U(N)$, 
which can be written in terms of the Omega factor as (\ref{DimR_Omega}).
Note that characters are class functions. 
The quantity (\ref{2pt_RHS_trace}), equivalently (\ref{bilinear_map_two_pt_single_matrix}),  
gives a bilinear map $W\otimes W\rightarrow \mathbb{C}$, 
where $W$ is a basis of $Z(\mathbb{C}[S_n])$, 
and we also find that it is non-degenerate i.e. it is invertible for $N>n$. 
We denote the nondegenerate bilinear form by $\eta$, with which 
we obtain a commutative Frobenius algebra. We denote the Frobenius algebra by 
$(Z(\mathbb{C}[S_n]),\eta)$.
The two-point function can be diagonalised using  
the represetation basis \cite{0111222}
\begin{eqnarray}
tr^{(r)}(\Omega_n P^R P^S)
=\frac{n!}{N^n}  d_R DimR\delta_{RS}.
\end{eqnarray}

\quad 

We next consider the two-matrix model. 
The number of holomorphic gauge invariant operators
built from $m$ $X$'s and $n$ $Y$'s 
is the same as the number of 
equivalence classes 
under $\sigma \sim  h\sigma h^{-1}$ in $S_{m+n}$, 
where $h \in S_m\times S_n$. 
In fact the 
holomorphic gauge invariant operators 
can be associated with the equivalence classes 
as 
\begin{align}
{\cal O}(\sigma)&=(X)_{i_1}^{i_{\sigma(1)}}
\cdots (X)_{i_m}^{i_{\sigma(m)}}(Y)_{i_{m+1}}^{i_{\sigma(m+1)}}
\cdots (Y)_{i_{m+n}}^{i_{\sigma(m+n)}}
\nonumber \\
&=
tr_{m+n}(\sigma X^{\otimes m}\otimes Y^{\otimes n} ),
\end{align}
which have the invariance
\begin{align}
{\cal O}(\sigma)={\cal O}(h \sigma h^{-1}), \quad h\in S_m\times S_n.
\end{align}
Two-point functions are 
\begin{eqnarray}
\langle {\cal O}(\sigma){\cal O}(\tau) \rangle =
tr^{(r)}(\Omega_{m+n} [\sigma]_H[\tau]_H), 
\end{eqnarray}
where $H=S_m\times S_n$. 
We now define
\begin{eqnarray}
\theta_{\sigma,\tau} :=
tr^{(r)}(\Omega_{m+n} [\sigma]_H[\tau]_H). 
\label{Frobenius_form_two-matrix}
\end{eqnarray}
Because this gives a non-degenerate pairing $W\otimes W \rightarrow \mathbb{C}$, 
where $W$ is the space spanned by elements in $\mathbb{C}[S_{m+n}]_H$, 
we can construct a Frobenius algebra with the bilinear form.  
As we emphasised in (\ref{noncommutative_restriction}), 
this algebra is noncommutative.
The bilinear form (\ref{Frobenius_form_two-matrix}) 
can be diagonalised as \cite{0801.2061}
\begin{eqnarray}
tr^{(r)}(\Omega_{m+n}P^R_{A,\mu\nu}P^S_{B,\rho\sigma})
=
d_A DimR \frac{m!n!}{N^{m+n}}\delta_{RS}
 \delta_{AB}
 \delta_{\mu \sigma}
 \delta_{\nu \rho}.
\end{eqnarray}

Recently singlet gauge invariant operators under the global $O(N_f)$ symmetry
were studied in \cite{1608.03188}. 
We can label the singlet operators in terms of some permutation as 
\begin{align}
{\cal S}(\sigma)&=
(\Phi_{a_1})_{i_1}^{i_{\sigma(1)}}
(\Phi_{a_1})_{i_2}^{i_{\sigma(2)}}
(\Phi_{a_2})_{i_3}^{i_{\sigma(3)}}
(\Phi_{a_2})_{i_4}^{i_{\sigma(4)}}
\cdots
(\Phi_{a_{n/2}})_{i_{n-1}}^{i_{\sigma(n-1)}}
(\Phi_{a_{n/2}})_{i_{n}}^{i_{\sigma(n)}}
\nonumber \\
&=
tr_{n}(\sigma \Phi_{a_1}\otimes \Phi_{a_1}\cdots \Phi_{a_{n/2}}\otimes \Phi_{a_{n/2}} ),
\end{align}
which are invariant under the conjugation
by the wreath product group $S_{n/2}[S_2]$
\begin{align}
{\cal S}(\sigma)={\cal S}(h \sigma h^{-1}), \quad h\in S_{n/2}[S_2].
\end{align}
The two-point functions are \cite{1608.03188}
\begin{align}
\langle {\cal S}(\sigma){\cal S}(\tau) \rangle &=
\frac{N^nN_f^{n/2}}{n!}
\varphi_{\sigma,\tau}
\nonumber \\
\varphi_{\sigma,\tau} &=
\frac{1}{N_f^{n/2}}
\sum_{\rho\in S_n}
W(\rho)tr^{(r)}(\Omega_{m+n} [\sigma]_H\rho [\tau]_H \rho^{-1}), 
\end{align}
where $H=S_{n/2}[S_2]$.
The $W(\rho)$ is a function 
of the form $N_f^{z(\rho)}$,
with $z(\rho)=z(h_1\rho h_2)$, $h_1,h_2\in H$. 
They can not be diagonalised by 
the representation basis in (\ref{expansion_restricted_operator})
but can be diagonalised by the basis 
in (\ref{representation_wreath_product}).
The two point functions 
can also be regarded as a map from 
$W\otimes W \rightarrow \mathbb{C}$,
where $W$ is a basis of $\mathbb{C}[S_{n}]_{S_{n/2}[S_2]}$.

\quad

It is very interesting 
that two-point functions are 
entirely 
expressed in terms of symmetric group quantities. 
A consequence of this fact might be an interpretation of 
the correlation functions in terms of 
two-dimensional topological field theories \cite{1301.1980}. 
Associating the two-point functions with the non-degenerate bilinear form
of the Frobenius algebra, 
we can interpret the two-point functions of the matrix model as 
two-point functions of the two-dimensional topological field theory.

If the algebras we are considering are commutative, 
they will describe two-dimensional field theories 
that may be 
associated with closed string theories. 
On the other hand, 
noncommutative algerbas may be relevant for open string theories. 
Because an one-loop non-planar 
open string diagram is topologically equivalent to a closed propagator, 
both open string algebras and closed string algebras 
should be described in an algebraic framework. 
We will clarify how open string degrees of freedom
and closed string degrees of freedom are described in the algebras 
we have discussed in this section.


\section{Simplest open-closed TFT}
\label{sec:simplest_open_closed_TFT}

In this section we shall review 
how 
open-closed topological systems are described by
semi-simple algebras \cite{9212154,9205031}.  
As a 
prototype of the relation between two-dimensional topological 
field theories and Frobenius algebras, we consider the group algebra of 
the symmetric group $S_n$. 
Let $\sigma_i$ be a basis of $\mathbb{C}[S_n]$, the group algebra of $S_n$. 
In order to make this algebra Frobenius, 
we now adopt the following bilinear form, 
\begin{eqnarray}
g_{ij}=C_{ik}{}^lC_{jl}{}^k,  
\label{bilinear_form_open}
\end{eqnarray}
where $C_{ij}{}^k$ is the structure constant of the algebra,
\begin{eqnarray}
\sigma_i\sigma_j=C_{ij}{}^k \sigma_k. 
\end{eqnarray}
This algebra is noncommutative
$C_{ij}{}^k\neq C_{ji}{}^k$.
We call this noncommutative Frobenius algebra $(\mathbb{C}[S_{n}],g)$. 
The bilinear form can also be written as 
\begin{eqnarray}
g_{ij}=tr^{(r)}(\sigma_i\sigma_j). 
\end{eqnarray}
The nondegenerate bilinear form is also called 
{\bf a Frobenius pairing}.

We now define {\bf the dual basis} by  
\begin{eqnarray}
\sigma^i=\frac{1}{n!}\sigma_i^{-1},  
\end{eqnarray}
and the inverse of $g_{ij}$ is given by  
\begin{eqnarray} 
g^{ij}=tr^{(r)}(\sigma^i\sigma^j).
\end{eqnarray}
Then we have 
\begin{eqnarray}
g_{ij}g^{jk}=g_{i}{}^k, \quad 
g_{i}{}^j=tr^{(r)}(\sigma_i\sigma^j)
\label{gg=g1}
\end{eqnarray}
and 
\begin{eqnarray}
g_{i}{}^j g_{j}{}^k=g_{i}{}^k. 
\label{gg=g2}
\end{eqnarray}
By these properties, the bilinear form is called non-degenerate in the basis. 
The $g_{i}{}^j$ is associated with the propagator of 
the two-dimensional theory. 
The property (\ref{gg=g2}) expresses the fact that the two-dimensional theory has  
the vanishing Hamiltonian, which is a general feature of 
topological quantum field theories \cite{Atiyah}.  
The dual basis also satisfies 
\begin{eqnarray}
\sigma^i=g^{ij}\sigma_j.
\label{dual_basisi_inverse_metric}
\end{eqnarray}
A comment about the relation between 
(\ref{bilinear_form_open}) 
and (\ref{dual_basisi_inverse_metric})
will be given in
appendix \ref{app:dual_basis_metric}.  
The difference between the basis and the dual basis corresponds to 
the different orientations of the boundaries.

The structure constant can be interpreted as the three-point vertex
of the theory, which has the form
\begin{align}
&C_{ij}{}^{k}=tr^{(r)}(\sigma_i\sigma_j\sigma^k) .
\label{algebra_sigma_i_g_C}
\end{align}
Reflecting the noncommutativity, 
the theory may be associated with 
two-dimensional surfaces corresponding to open string worldsheets. 
We then regard the Frobenius algebra as describing 
a topological open string theory. 
We also call the bilinear form {\bf an open string metric}.
Indices are raised or lowered by the open string metric.

\quad

\begin{figure}
\centering
\includegraphics[clip,scale=0.6]{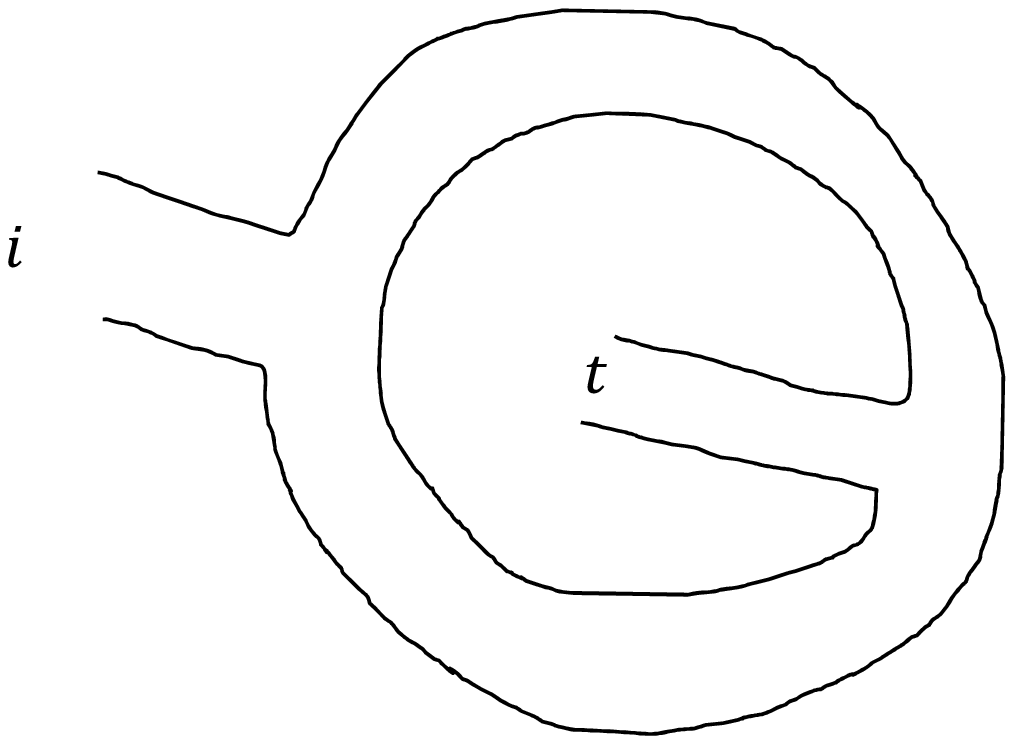}
\caption{The RHS of (\ref{CC=eta}), $C_{ik}{}^{l}C_{lt}{}^{k}$, is shown in this figure. } 
\label{fig:open_one_loop}
\end{figure}

Let us next explain how the dual closed theory arises 
in this description. 
The dual closed string propagator is given by computing  
the open string one-loop non-planar diagram (see figure 
\ref{fig:open_one_loop}), 
\begin{eqnarray}
\eta_{it}=C_{ik}{}^{l}C_{lt}{}^{k}.
\label{CC=eta}
\end{eqnarray}
We call it {\bf a closed string metric}.
From (\ref{algebra_sigma_i_g_C}) with the orthogonality relation of 
representations (\ref{orthogonality_representations}), we also have 
\begin{eqnarray}
\eta_{ij}=tr^{(r)}([\sigma_i][\sigma_j]),
\label{eta_[sigma][sigma]}
\end{eqnarray}
where $[\sigma]$ is defined in (\ref{sum_conjugacy_class}). 
We can easily check that 
the closed string propagator is a projection operator
\begin{eqnarray}
\eta_{i}{}^j\eta_j{}^k=\eta_i{}^k,
\end{eqnarray}
where indices are raised by the open string metric,
\begin{eqnarray}
\eta_i{}^j=g^{jk}\eta_{ik}.
\end{eqnarray}
The open string metric (\ref{bilinear_form_open}) 
secures 
the topological property of the closed string theory. 
In this way we obtain a topological closed string theory from the 
topological open string theory.
The topological closed string theory is described by the centre of $\mathbb{C}[S_n]$ with 
the bilinear form $\eta_{ij}$. 
This Frobenius algebra is indeed what we have introduced in the previous section, 
which we denoted by  
($Z(\mathbb{C}[S_n])$, $\eta$), 
if we ignore the $\Omega_n$ by taking a large $N$ limit.\footnote{
Because the $\Omega_n$ has an $1/N$ expansion as in (\ref{Omega_factor}), 
by taking a large $N$ limit, we ignore the contribution of $\Omega_n$ 
in (\ref{twopt_single_matrix}).
} 

\begin{figure}
\centering
\includegraphics[clip,scale=0.6]{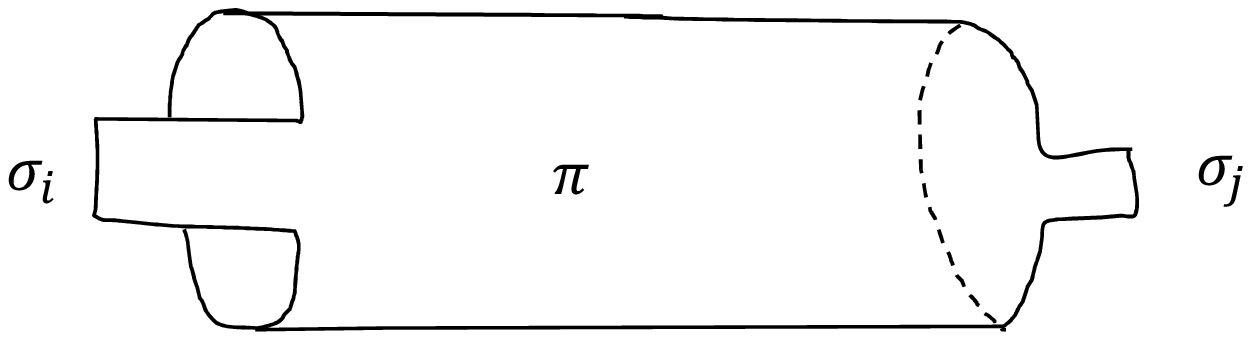}
\caption{equaiton (\ref{closed_propagation_in_terms_of_pi}) is graphycally shown} 
\label{fig:closed_propagation_pi}
\end{figure}

For convenience, let us introduce a map from 
$\mathbb{C}[S_n]$ to $Z(\mathbb{C}[S_n])$ by 
\begin{eqnarray}
\pi(\sigma)=\sum_i \sigma_i \sigma \sigma^i
=[\sigma]. 
\label{open_closed_duality}
\end{eqnarray}
From this definition it follows $\pi^2=\pi$. 
This gives a map from 
the open string states to 
the closed string states \cite{0609042}. 
We can also define the map in terms of the 
the closed string propagator as
\begin{eqnarray}
\eta_i{}^j\sigma_j=[\sigma_i].
\end{eqnarray}
Here we stress that the $\eta$ is a projector onto the centre 
of the group algebra $\mathbb{C}[S_n]$,
\begin{eqnarray}
\eta_i{}^j[\sigma_j]=[\sigma_i]. 
\end{eqnarray}
Using the map we have  
\begin{eqnarray}
\eta_{ij}=tr^{(r)}(\pi(\sigma_i)\pi(\sigma_j))
=tr^{(r)}(\sigma_i \pi(\sigma_j)), 
\label{closed_propagation_in_terms_of_pi}
\end{eqnarray}
where 
$\pi$ looks like a closed string mode propagating between two open string states
(see figure \ref{fig:closed_propagation_pi}). 
The property 
$\eta^2=\eta$ can be derived from $\pi^2=\pi$. 
The action of $\pi$ is diagonalised by the representation basis
\begin{align}
\pi(P^R)=P^R. 
\end{align}
Closed string states are labelled by a Young diagram.

Starting with the noncommutative Frobenius algebra $(\mathbb{C}[S_{n}],g)$, 
the commutative Frobenius algebra also 
$(Z(\mathbb{C}[S_{n}]),\eta)$
comes in the game. 
We call this theory an open-closed topological theory. 
An axiomatic description of open-closed topological field 
theories was given in \cite{Segal,0609042}.

We have explicitly reviewed the topological theory description of 
the group algebra obtained from
the symmetric group, and the construction 
works for any semi-simple algebras \cite{9205031,9212154}. 
In \cite{1403.6572} we have discussed topological field theories 
obtained from walled Brauer algebras.


\section{Frobenius algebras with restricted structure}
\label{sec:restricted_algebra}

In the previous section we have reviewed that 
starting with a noncommutative Frobenius algebra 
with the Frobenius pairing (\ref{bilinear_form_open}), 
a commutative Frobenius algebra is uniquely obtained  
via (\ref{CC=eta}), which is considered as
the open-closed duality.
In this section we will study Frobenius algebras associated 
with the algebras $\mathbb{C}[S_n]_H$ 
in more detail. Because these algebras are noncommutative, 
we are tempted to interpret these theories as open string theories. 
We will explore some aspects of the theories from the view point of 
the open-closed duality.

Recall that a basis of the algebra $\mathbb{C}[S_n]_H$ is spanned by 
\begin{align}
[\sigma]_H=\frac{1}{m!n!}\sum_{h\in H} h\sigma h^{-1}, 
\end{align}
which satisfy 
(\ref{commuting_algebra_restriction_H})-(\ref{restricted_sum_inverse}). 
Let $\Xi_{ij}{}^k$ be the structure constant of the algebra, 
\begin{eqnarray}
[\sigma_i]_H[\sigma_j]_H=\Xi_{ij}{}^k[\sigma_k]_H. 
\end{eqnarray}
This is a noncommutative algebra
\begin{align}
\Xi_{ij}{}^k\neq \Xi_{ji}{}^k. 
\end{align}

In general the
there are several choices for the Frobenius pairing.
As we discussed in the last section, 
one possible Frobenius pairing for the noncommutative 
algebra is to use the prescription 
(\ref{bilinear_form_open}) as 
\begin{align}
G_{ij}=\Xi_{ik}{}^l \Xi_{jl}{}^{k}. 
\label{def_of_G_ij}
\end{align}
We call 
this noncommutative Frobenius algebra ($\mathbb{C}[S_{m+n}]_H$, $G$). 
The open-closed duality (\ref{CC=eta}) 
gives a topological closed string theory 
described by the commutative Frobenius algebra. 
We construct it explicitly in 
appendix \ref{TFT_from_[sigma]_H}.

On the other hand,
it is also natural for us to use the two-point function as a Frobenius pairing of 
$\mathbb{C}[S_{m+n}]_{H}$. 
For example when $H=S\times S_n$, it is given by 
\begin{align}
\theta_{ij}=tr^{(r)}([\sigma_i]_H [\sigma_j]_H),
\end{align} 
where we have ignored $\Omega_{m+n}$ 
to make this discussion simple. 
We will interpret it as an open string propagator.  
(It may also be interpreted as a cylinder amplitude by considering a triangulation 
with some links restricted in the subgroup \cite{1301.1980,1403.6572}.)
In section \ref{sec:review_matrix_model}
this Frobenius algebra was called ($\mathbb{C}[S_{m+n}]_H$, $\theta$). 
A difference 
between these two Frobenius algebras, 
($\mathbb{C}[S_{m+n}]_H$, $G$) and ($\mathbb{C}[S_{m+n}]_H$, $\theta$), 
is that 
$G_{ij}$ is written in terms of $\Xi$ as (\ref{def_of_G_ij}), 
while 
$\theta_{ij}$ does not  
satisfy $\Xi_{ik}{}^l \Xi_{jl}{}^{k}=\theta_{ij}$\footnote{
We find 
\begin{align}
&
\theta_{i}{}^l=tr^{(r)}([\sigma_i]_H[\sigma^l]_H)
\nonumber \\
&
\Xi_{ik}{}^l \Xi_{jl}{}^{k}=tr^{(r)}([\sigma_k]_H[\sigma^k]_H[\sigma_j]_H[\sigma_i]_H),
\label{G1B2open_correlator_planar}
\end{align}
where
\begin{eqnarray}
\sum_i [\sigma_i]_H[\sigma^i]_H=\sum_{R,A}\frac{M^R_A}{d_Rd_A}P^R_A. 
\label{phi^1phi_1_theta}
\end{eqnarray}
This is consistent with 
appendix \ref{app:dual_basis_metric}. 
}.
This means that 
the propagator of 
the closed string theory dual to 
($\mathbb{C}[S_{m+n}]_H$, $\theta$) does not have to be a projector. 
The relation between $G$ and $\theta$ will be shown in appendix \ref{TFT_from_[sigma]_H}. 

When $H=S_{n/2}[S_2]$, 
Frobenius algebra  
($\mathbb{C}[S_{m+n}]_H$, $\varphi$)
can be naturally considered, 
where 
\begin{align}
\varphi_{ij}=\frac{1}{N_f^{n/2}}\sum_{\rho\in S_n}W(\rho)
tr^{(r)}([\sigma_i]_H \rho [\sigma_j]_H\rho^{-1}).
\end{align}

\subsection{Noncommutative 
Frobenius algebra ($\mathbb{C}[S_{m+n}]_{S_m\times S_n}$, $\theta$)}

In this subsection we consider the Frobenius algebra 
$(\mathbb{C}[S_{m+n}]_{H}$, $\theta$) for
$H=S_m\times S_n$. 

The indices of the open string metric $\theta_{ij}$ can be raised by 
$g^{ij}$ \cite{1403.6572} as
\begin{align}
\theta_{i}{}^{j}=tr^{(r)}([\sigma_i]_H [\sigma^j]_H), \quad 
\theta^{ij}=tr^{(r)}([\sigma^i]_H [\sigma^j]_H),
\end{align} 
which satisfy $\theta_{ij}\theta^{jk}=\theta_i{}^k$.
It has another expression 
\begin{align}
\theta_{i}{}^j=\sum_{R,A,\mu\nu}\frac{d^R}{d_A}
\chi^R_{A,\mu\nu}(\sigma_i)\chi^R_{A,\nu\mu}(\sigma^j).
\end{align} 
Note that $\sum_i\theta_i{}^i=\sum_{R,A}(M^R_A)^2$ counts the dimension of the vector space, 
i.e. the number of independent operators expressed by the form $P^R_{A,\mu\nu}$. 
Some partition functions of this theory were computed in \cite{1403.6572} 
with a different interpretation from here.

The dual basis is defined in terms of the open string metric as  
\begin{eqnarray}
 [\sigma^k]_H=\theta^{kl}[\sigma_l]_H,
\end{eqnarray}
which is also the same as 
\begin{eqnarray}
 [\sigma^k]_H=[g^{kl}\sigma_l]_H.
\end{eqnarray}
Closed string operators can be defined 
by generalising (\ref{open_closed_duality}) to
\begin{eqnarray}
\Pi([\sigma]_H)
:=\sum_k [\sigma_k]_H[\sigma]_H [\sigma^k]_H.
\label{closed_restricted}
\end{eqnarray}
(Another way of defining closed string operators 
is to use (\ref{map_open_to_closed}).)
They belong to the centre $Z(\mathbb{C}[S_{m+n}]_H)$,\footnote{
For an associative algebra, 
$\pi(\phi)=\phi_i \phi \phi^i$ gives a central element, 
where $\phi_i$ is a basis and $\phi^i$ is the dual basis \cite{0609042}. 
Let $C_{ij}{}^k$ be the structure constants. 
Then we can show the following
\begin{eqnarray}
\phi_j \pi(\phi)=
\phi_j (\phi_i\phi\phi^i)=C_{ji}{}^k\phi_k\phi\phi^i
=\phi_k\phi( C^k{}_{ji}\phi^i)=(\phi_k\phi\phi^k) \phi_j=
\pi(\phi)\phi_j. 
\end{eqnarray}
}
\begin{eqnarray}
[\sigma]_H\Pi([\tau]_H)=\Pi([\tau]_H)[\sigma]_H .
\end{eqnarray}
This property is manifest in the form
\begin{eqnarray}
\Pi([\sigma]_H)
=\sum_{R,A}
\frac{1}{(d_A)^2d_R}\chi^R_A(\sigma)P^R_A.
\label{center_expand_basis}
\end{eqnarray}
It is found that 
$\Pi^2$ is not equal to $\Pi$ due to an extra factor in the following, 
but it is still in the centre
\begin{eqnarray}
\Pi(\Pi([\sigma_i]_H))=
\sum_{R,A}
\frac{e^{t_{R,A}}}{(d_A)^2 d_R}\chi^R_A(\sigma_i)P^R_A.
\end{eqnarray}
where 
we have defined for our convenience
\begin{eqnarray}
e^{t_{R,A}}:=\frac{M^R_A}{d_Ad_R}.
\label{def_t_RA}
\end{eqnarray}
It will be regarded as an area of a closed string cylinder. 
Inverting (\ref{center_expand_basis}), 
central elements in the representation basis 
can be expanded as 
\begin{align}
P^R_A=
&\frac{d_R}{(m+n)!}e^{-t_{R,A}}\sum_{\sigma\in S_{m+n}}
\chi^R_A(\sigma^{-1})\Pi([\sigma]_H)
\cr
=&
d_R e^{-t_{R,A}}\sum_{i}
\chi^R_A(\sigma^{i})\Pi([\sigma_i]_H). 
\end{align}


We now compute the closed string propagator via the open-closed duality as 
\begin{eqnarray}
Z_{ij}=\Xi_{ik}{}^l\Xi_{lj}{}^k
=
tr^{(r)}([\sigma_j]_H\Pi([\sigma_i]_H)).
\label{closed_propagator_restricted}
\end{eqnarray}
It also takes the following form \cite{1403.6572}
\begin{eqnarray}
Z_{ij}=\sum_{R,A,\mu,\nu}
\frac{1}{(d_A)^2}\chi^R_{A,\mu\mu}(\sigma_i)\chi^R_{A,\nu\nu}(\sigma_j).
\end{eqnarray}
Using the open string metric $\theta$, 
we define
\begin{eqnarray}
Z_i{}^l=Z_{ij}\theta^{jl}=tr^{(r)}([\sigma^l]_H\Pi([\sigma_i]_H)), 
\end{eqnarray}
and 
$Z^{ij}$ is defined by $Z^{ij}=\theta^{ik}\theta^{jl}Z_{kl}$. 
These are consistent with 
$Z_{ij}Z^{jk}=Z_i{}^k$, but $Z_i{}^k$ does not satisfy $Z_i{}^lZ_l{}^j=Z_i{}^j$. In fact it can be shown that 
\begin{eqnarray}
Z_i{}^lZ_l{}^j=tr^{(r)}(\Pi([\sigma_i]_H)\Pi([\sigma^j]_H))
=tr^{(r)}([\sigma_i]_H\Pi^2([\sigma^j]_H)),
\end{eqnarray}
which is not equal to $Z_i{}^j$ because of  $\Pi^2\neq \Pi$.

Closed string states can be generated from open string states by acting with 
$Z_i{}^l$, which is a map onto the centre of the algebra,
\begin{eqnarray}
Z_i{}^l[\sigma_l]_H=\Pi([\sigma_i]_H).
\label{map_open_to_closed}
\end{eqnarray}
%
%
%
The $\Pi$ is diagonalised by the representation basis, 
\begin{eqnarray}
\Pi(P^R_{A,\mu\nu})
=\delta_{\mu\nu}
\frac{1}{d_A d_R}P^R_{A}.
\end{eqnarray}
In particular the eigenvalue of $\Pi^n:=(\Pi)^n$ is given by 
\begin{eqnarray}
\Pi^n (P^R_A)
=e^{n t_{RA}} P^R_A. 
\label{Pi^2_on_PRA}
\end{eqnarray}
Here $t_{R,A}$ might be considered to be a parameter to measure
the area of 
a closed string propagator \cite{Segal}. 
The area depends on the set of representations $(R,A)$. 
Some partition functions of this closed string theory can be computed.
For example, a torus amplitude is given by\footnote{
If we use the open string metric given in 
(\ref{Open_metric_topological_restriction}), 
\begin{eqnarray}
\tilde{Z}_i{}^l=G^{lj}Z_{ij}
\end{eqnarray}
satisfies $\tilde{Z}^2=\tilde{Z}$. 
The trace part of $\tilde{Z}_i{}^l$ counts the number of $P^R_A$, 
as in (\ref{dimension_central_[sigma]_H}).
} 
\begin{eqnarray}
Z_{ij}\theta^{ij}
=\sum_{R,A}e^{t_{R,A}}, 
\end{eqnarray}
and another torus can be computed by 
\begin{eqnarray}
Z_{ij}Z^{ij}=\sum_{R,A}(e^{t_{R,A}})^2.
\end{eqnarray}

Some representations give the zero eigenvalue $t_{R,A}=0$. 
Irreducible representations of $H=S_{m}\times S_{n}$ are 
labelled by a set of two Young diagrams 
$A=(\alpha,\beta)$, where 
$\alpha$ is a Young diagram with $m$ boxes 
and $\beta$ is a Young diagram with $n$ boxes.
The number of times the $A$ appears in $R$ is given by 
the Littlewood-Richardson coefficient,  
$M^R_A=g(\alpha,\beta;R)$.\footnote{
The Littlewood-Richardson coefficients are defined by 
\begin{eqnarray}
g(\alpha,\beta;R)=
\frac{1}{m!n!}\sum_{\sigma_1\in S_{m},\sigma_2\in S_{n}}
\chi_{\alpha}(\sigma_1^{-1})\chi_{\beta}(\sigma_2^{-1})
\chi_R(\sigma_1\circ \sigma_2 ).
\end{eqnarray}
} 
When both $R$ and $A$ are  
the symmetric representations $R=[m+n]$, $A=([m],[n])$, or 
both $R$ and $A$ are  
the anti-symmetric representations
$R=[1^{m+n}]$, $A=([1^m],[1^n])$,\footnote{
The Young diagram of $[p]$ has $p$ boxes in the first row, 
while the Young diagram of $[1^p]$ has $p$ boxes in the first column. 
} we have an zero eigenvalue $t_{R,A}=0$.
But for most representations we have $t_{R,A}\neq 0$.


Based on Brauer algebras open-closed systems 
can also be constructed very similarly following \cite{1403.6572}.

\subsection{Noncommutative 
Frobenius algebra ($\mathbb{C}[S_{n}]_{S_{n/2}[S_2]}$, $\varphi$)}

For the 
noncommutative algebra $\mathbb{C}[S_{n}]_{S_{n/2}[S_2]}$, 
as we discussed in section \ref{sec:review_matrix_model},
it is interesting to consider the following bilinear form,     
\begin{align}
&\varphi_{ij}=\frac{1}{N_f^{n/2}}\sum_{\rho\in S_n}W(\rho)
tr^{(r)}([\sigma_i]_H \rho [\sigma_j]_H\rho^{-1})
\nonumber \\
&\varphi^{ij}=\frac{1}{N_f^{n/2}}\sum_{\rho\in S_n}W(\rho)
tr^{(r)}([\sigma^i]_H \rho [\sigma^j]_H\rho^{-1})
\nonumber \\
&\varphi_{i}{}^{j}=\frac{1}{N_f^{n/2}}\sum_{\rho\in S_n}W(\rho)
tr^{(r)}([\sigma_i]_H \rho [\sigma^j]_H\rho^{-1}),
\end{align} 
which satisfy $\varphi_{ij}\varphi^{jk}=\varphi_{i}{}^k$.
The commutative algebra arising from the centre of 
the noncommutative algebra $\mathbb{C}[S_{n}]_{S_{n/2}[S_2]}$ 
can be associated with a topological closed string theory 
in the same manner as the last subsection.


\section{Summary}
\label{summary}

Based on the relationship between 
Frobenius algebras and two-dimensional topological field theories, 
we have 
discussed some aspects of the open-closed duality 
for Frobenius algebras that are  
related to the description of gauge invariant operators 
of ${\cal N}=4$ Super Yang-Mills.

We have focused on the 
the group algeba of symmetric group over the complex numbers.
Considering the conjugacy classes leads to   
an associative algeba spanned by the central elements, which 
can be associated with gauge invariant operators built from 
one kind of complex scalar in ${\cal N}=4$ Super Yang-Mills.
An interesting class of algebras can be introduced by considering 
equivalence relations using elements in subgroups of the symmetric group. 
The class of algebras were originally introduced 
to label gauge invariant operators 
in more general sectors 
than the one-complex scalar sector
in ${\cal N}=4$ Super Yang-Mills. 
These algebras are associative and noncommutative. 
These noncommutative algebras 
can be interpreted as 
topological open string theories 
based on the equivalence between 
Frobenius algebras and two-dimensional 
topological field theories.
We have associated  
two-point functions of ${\cal N}=4$ Super Yang-Mills with
the bilinear form of the Frobenius algebras. 
It has been concretely observed that 
closed string degrees of freedom come out from 
the centre of the noncommutative algebras.
The closed string states are labelled by a set of 
representations. 
It would be interesting to consider how 
the representations are interpreted in this context.

\qquad 

\noindent
{\bf Acknowledgements}

This work was supported by JSPS KAKENHI Grant Number 15K17673. 
We acknowledge useful conversations with 
Pablo Diaz, 
Robert de Mello Koch, 
Hai Lin, 
Sanjaye Ramgoolam,  
Ryo Suzuki 
at the ESF and STFC supported 
 workshop ``Permutations and Gauge String duality'' in 2014.


\appendix 
\renewcommand{\theequation}
{\Alph{section}.\arabic{equation}}

\section{Useful formulas}
\label{appendix_formulas}
\setcounter{equation}{0}

Let $D^R_{ij}(\sigma)$ be a representation matrix of $\sigma \in S_n$. 
The orthogonality relation is given by 
\begin{align}
\frac{1}{n!}\sum_{\sigma\in S_n}
D^R_{ij}(\sigma)
D^S_{kl}(\sigma^{-1})
=\frac{1}{d_R}\delta_{il}\delta_{jk}\delta_{RS},
\label{orthogonality_representations}
\end{align}
where $d_R$ is the dimension of an irreducible representation $R$ of $S_n$.

The dimension of an irreducible representation $R$ of $U(N)$ can be 
expressed in terms of the Omega factor  
(\ref{Omega_factor}) as 
\begin{align}
Dim R=\frac{N^n}{n!}\chi_R (\Omega_n),
\label{DimR_Omega}
\end{align}
where $\chi_R$ is the character of a representation $R$.



\section{Dual basis and the bilinear form}
\setcounter{equation}{0}
\label{app:dual_basis_metric} 

In this section, we give a supplementary explanation about 
the relationship between the bilinear form and the dual basis.

In order to make this discussion general,  
consider an associative algebra, where we denote elements of the algebra by $\phi_i$, 
the structure constants by $C_{ij}{}^k$. 
Here we will study the implication of the bilinear form determined by
\cite{9205031,9212154} 
\begin{eqnarray}
g_{ij}=C_{ik}{}^lC_{jl}{}^{k}.
\label{CC=g_app}
\end{eqnarray}
This makes the associative algebra Frobenius. 
The right-hand side represents 
the planar one-loop two-point function with the assignment of  
a 3pt-vertex to the structure constant. 




Multiplying the equation by $\phi^j$, 
the right-hand side gives
\begin{eqnarray}
C_{ik}{}^lC_{jl}{}^{k}\phi^j=C_{ik}{}^l\phi_l\phi^k=\phi_i\phi_k\phi^k,
\end{eqnarray}
while the left-hand side gives
\begin{eqnarray}
g_{ij}\phi^j=\phi_i.
\end{eqnarray}
We then obtain 
\begin{eqnarray}
\sum_k \phi_k\phi^k=1.
\label{def_dual_basis_ap2}
\end{eqnarray}
The condition (\ref{CC=g_app}) 
is equivalent to choosing the dual basis satisfying 
(\ref{def_dual_basis_ap2}). 

The Frobenius algebras appeared in section \ref{sec:simplest_open_closed_TFT} and 
Appendix \ref{TFT_from_[sigma]_H} 
have the Frobenius form 
in (\ref{CC=g_app}). 
On the other hand, 
the bilinear form in 
the Frobenius algebras introduced in section \ref{sec:restricted_algebra}
do not take the form (\ref{CC=g_app}). 

We can find the following equation, 
\begin{eqnarray}
C_{ij}{}^kg_{k0}=g_{ij}.
\label{frobenius_form_Cg=g}
\end{eqnarray}
This is always satisfied, and it is 
not a condition which determines the form of the metric.
This just gives a consistency condition, while (\ref{CC=g_app}) 
is a non-trivial constraint on the form of the metric.



\section{Another Frobenius algebra from $\mathbb{C}[S_{m+n}]_H$}
\setcounter{equation}{0}
\label{TFT_from_[sigma]_H}

In this appendix, 
the noncommutative Frobenius algebra $(\mathbb{C}[S_{m+n}]_H,G)$ will be studied, 
where $G$ is given in (\ref{def_of_G_ij}). 
We explicitly construct the commutative Frobenius algebra 
obtained from this noncommutative Frobenius algebra.  

We now have the open string metric given by 
\begin{eqnarray}
G_{ij}=\Xi_{ik}{}^l\Xi_{jl}{}^k=\sum_{R,A,\mu,\nu}
\frac{M^R_A}{(d_A)^2}\chi^R_{A,\mu\nu}(\sigma_i)\chi^R_{A,\nu\mu}(\sigma_j), 
\label{Open_metric_topological_restriction}
\end{eqnarray}
and we also have 
\begin{eqnarray}
&&G^{ij}=\sum_{R,A,\mu,\nu}
\frac{M^R_A}{(d_A)^2}\chi^R_{A,\mu\nu}(\tilde{\sigma}^i)
\chi^R_{A,\nu\mu}(\tilde{\sigma}^j),
\nonumber \\
&&G_i{}^{j}=\sum_{R,A,\mu,\nu}
\frac{M^R_A}{(d_A)^2}\chi^R_{A,\mu\nu}(\sigma_i)\chi^R_{A,\nu\mu}(\tilde{\sigma}^j),
\end{eqnarray}
where 
the new dual basis has been introduced by   
\begin{eqnarray}
\tilde{\sigma}^i=G^{ij}\sigma_j.
\end{eqnarray}
They satisfy
$G_{ij}G^{jk}=G_i{}^k$ and 
$G_i{}^jG_j{}^k=G_i{}^k$, 
and 
$\Xi_{ij}{}^k G_{k0}=G_{ij}$.
The two dual bases are simply related by 
\begin{eqnarray}
\chi^R_{A,\mu\nu}(\tilde{\sigma}^i)=e^{-t_{R,A}}\chi^R_{A,\mu\nu}(\sigma^i),
\end{eqnarray}
where we recall that $t_{R,A}$ was defined in (\ref{def_t_RA}).
As is expected, we have 
\begin{eqnarray}
G_i{}^{j}\sigma_j=[\sigma_i]_H.
\end{eqnarray}

A closed string propagator was obtained in (\ref{closed_propagator_restricted}). 
Because 
indices of $Z_{ij}$ are raised by the open string metric $G^{ij}$, we introduce
\begin{eqnarray}
\tilde{Z}_{i}{}^j:=Z_{ik}G^{kj}=
\sum_{R,A,\mu,\nu}
\frac{1}{(d_A)^2}\chi^R_{A,\mu\mu}(\sigma_i)\chi^R_{A,\nu\nu}(\tilde{\sigma}^j).
\end{eqnarray}
We can show 
$\tilde{Z}_{i}{}^j\tilde{Z}_{j}{}^k=\tilde{Z}_{i}{}^k$ as expected. 
Another consistency check is  
$\tilde{Z}_{ij}\tilde{Z}^{jk}=\tilde{Z}_i{}^k$, where 
\begin{eqnarray}
\tilde{Z}^{ij}
:=Z_{kl}G^{ki}G^{lj}
=\sum_{R,A,\mu,\nu}
\frac{1}{(d_A)^2}\chi^R_{A,\mu\mu}(\tilde{\sigma}^i)\chi^R_{A,\nu\nu}(\tilde{\sigma}^j).
\end{eqnarray}

Similarly to (\ref{open_closed_duality}) and (\ref{closed_restricted}), 
closed string states are defined by 
\begin{align}
\tilde{\Pi}(\sigma_i)=
[\sigma_k]_H[\sigma_i]_H[\tilde{\sigma}^k]_H
&=\sum_{R,A}\frac{1}{M^R_A d_A}\chi^R_A(\sigma_i)P^R_A
\end{align}
or 
\begin{align}
\tilde{\Pi}(\sigma_i)=\tilde{Z}_i{}^j[\sigma_j]_H.
\end{align}
The inverse transformation gives the expansion of 
central projectors 
in $Z(\mathbb{C}[S_n]_H)$ as 
\begin{eqnarray}
P^R_A=d_R
\sum_i
\chi^R_A(\sigma^i)\tilde{\Pi}(\sigma_i) .
\end{eqnarray}
We find that $\tilde{\Pi}^2=\tilde{\Pi}$, and  
on the representation basis $\tilde{\Pi}$ is diagonalised by  
\begin{eqnarray}
\tilde{\Pi}(P^R_A)=
P^R_A.
\end{eqnarray}

From appendix \ref{app:dual_basis_metric}, 
we expect the dual basis introduced in this section should 
satisfies $\sum_i[\sigma_i]_H[\tilde{\sigma}^i]_H=1$, and it is the case: 
\begin{align}
&\sum_i[\sigma_i]_H[\tilde{\sigma}^i]_H
\nonumber \\
=&
\sum_{i,j}[\sigma_i]_H [\sigma_j]_H G^{ij}
\nonumber \\
=&
\sum_{R,A,\mu\nu} \frac{1}{d_A}\chi^R_{A,\mu\nu}(\sigma_i)
P^R_{A,\mu\nu}
\sum_{S,B,\alpha\beta} \frac{1}{d_B}
\frac{d_Sd_B}{M^S_B}
\chi^S_{B,\alpha\beta}(\sigma^i)
P^S_{B,\alpha\beta}
\nonumber \\
=&
\sum_{R,A}P^R_A=1.
\end{align}

\quad 

The trace part of $\tilde{Z}_i^{j}$ has a mathematical meaning. 
Because $\tilde{Z}_i^{j}$ is a projection operator onto the centre 
$Z(\mathbb{C}(S_n)_H)$, 
the trace part counts the dimension of the vector space, 
\begin{eqnarray}
\tilde{Z}_{i}{}^i=\sum_{R,A}\frac{M_{RA}}{M_{RA}}
=\sum_{R,A}\delta_{M_{R,A},0}
\label{dimension_central_[sigma]_H}.
\end{eqnarray}
Indeed this coincides the number of the operators $P^R_A$.

\section{Another representation basis}
\label{TFT_covariant_basis}

In this appendix, we will study Frobenius algebra $(\mathbb{C}[S_n]_H,\theta)$
in terms of a different representation basis introduced in \cite{0711.0176,1301.1980}.

Let us introduce branching coefficients
$B^{\Lambda\rightarrow 1_H,\alpha }_{l}$
of an irreducible representation $\Lambda$ 
of 
$S_n$, where $l$ runs over $1,\cdots,d_{\Lambda}$. 
They are the quantities satisfying 
\begin{eqnarray}
\sum_{\alpha}B^{\Lambda\rightarrow 1_H,\alpha }_{l_1}
B^{\Lambda\rightarrow 1_H,\alpha }_{l_2}
=
\frac{1}{|H|}
\sum_{h\in H}D^{\Lambda}_{l_1l_2}( h)
\end{eqnarray}
and 
\begin{eqnarray}
\sum_{l}B^{\Lambda\rightarrow 1_H,\alpha }_{l}
B^{\Lambda\rightarrow 1_H,\beta}_{l}
=\delta_{\alpha\beta}.
\end{eqnarray}
Here 
the indices $\alpha,\beta$ run over $1,\cdots,M^{\Lambda}_{1_H}$, 
where 
$M^{\Lambda}_{1_H}$ is the number of times the singlet representation 
$1_H$ of $H$ appears in $\Lambda$. 
From the reality of the representation of $S_n$, 
we have $B^{\dagger}=B$. 

Another important quantities are Clebsh-Gordan coefficients
$S^{RS,\Lambda\tau}_{ab,l}$. Here 
$R,S,\Lambda$ are irreducible representations of $S_n$, and 
$a,b$ and $l$ run over $1,\cdots,d_{R}$, $1,\cdots,d_{S}$ and 
$1,\cdots,d_{\Lambda}$, respectively.
$\tau$ is a multiplicity label counting the number of times 
$\Lambda$ appears in $R\otimes R$, running over 
$1,\cdots,C(R,S,\Lambda)$.\footnote{
We introduce
\begin{eqnarray}
C(R,S,\Lambda)=
\frac{1}{n!}\sum_{\sigma\in S_n}\chi_R(\sigma)\chi_S(\sigma)\chi_{\Lambda}(\sigma).
\end{eqnarray}
} 
For our convenience, we define
\begin{eqnarray}
\tilde{S}^{RR,\Lambda\tau,\alpha}_{ab}
:=\sum_{l}
S^{RR,\Lambda\tau}_{ab,l}B^{\Lambda\rightarrow 1_H,\alpha }_{l}.
\label{def_tildeS}
\end{eqnarray}

In terms of these group theoretical quantities, 
we can show\footnote{
In (\ref{expand_sigma_H_CG_basis}), the LHS is invariant under 
under $\sigma \rightarrow h\sigma h^{-1}$, where $h\in H$. 
Correspondingly the RHS should also be invariant under this transformation, 
which can be checked as follows,
\begin{align}
&
D^R_{ab}(h\sigma^i h^{-1})
\tilde{S}^{RR,\Lambda\tau,\alpha}_{ab}
\nonumber \\
=&
D^R_{ac}(h)D^R_{cd}(\sigma^i) D^R_{bd}(h)
S^{RR,\Lambda\tau}_{ab,l}B^{\Lambda\rightarrow 1_H,\alpha}_l
\nonumber \\
=&
D^R_{cd}(\sigma^i) D^{\Lambda}_{ml}(h^{-1})
S^{RR,\Lambda\tau}_{cd,m}B^{\Lambda\rightarrow 1_H,\alpha}_l
\nonumber \\
=&
D^R_{cd}(\sigma^i) 
S^{RR,\Lambda\tau}_{cd,m}B^{\Lambda\rightarrow 1_H,\alpha}_m,
\end{align}
where from the second line to the third line we have used \cite{Hamermesh} 
\begin{align}
D^R_{ac}(h)D^R_{bd}(h)S^{RR,\Lambda\tau}_{ab,l}
=D^R_{lm}(h)S^{RR,\Lambda\tau}_{cd,m},
\end{align}
and the following equation has been used to obtain the last line, 
\begin{eqnarray}
\sum_l D^{\Lambda}_{ml}(h^{-1})
B^{\Lambda\rightarrow 1_H,\alpha}_l
%
=B^{\Lambda\rightarrow 1_H,\alpha}_m.
\end{eqnarray}
} 
\begin{eqnarray}
[\sigma_i]_H=
\sum_{R,\Lambda,\tau,\alpha,l} 
d_R 
D^R_{ab}(\sigma_i)
\tilde{S}^{RR,\Lambda\tau,\alpha}_{ab}
P^{R,\Lambda,\tau,\alpha},
\label{expand_sigma_H_CG_basis}
\end{eqnarray}
where
\begin{align}
P^{R,\Lambda,\tau,\alpha}
&=\sum_iD^R_{ba}(\sigma^i)
\tilde{S}^{RR,\Lambda\tau,\alpha}_{ab}\sigma_i
\\
&=\sum_iD^R_{ba}(\sigma^i)
\tilde{S}^{RR,\Lambda\tau,\alpha}_{ab}[\sigma_i]_H.
\label{rep_basis_P_R_Lambda_tau}
\end{align}

When $H=S_{n/2}[S_2]$, the Branching coefficients are zero if $\Lambda$ is odd. 
If $\Lambda$ is even, $M^{\Lambda}_{1_H}=1$, so 
the index $\alpha$ is trivial;
\begin{align}
P^{R,\Lambda,\tau}
&=\sum_iD^R_{ba}(\sigma^i)
S^{RR,\Lambda\tau}_{ab,l}B_l^{\Lambda\rightarrow 1_H}[\sigma_i]_H.
\label{representation_wreath_product}
\end{align}
This property has played an important role in \cite{1608.03188}. 

\quad 

For later convenience 
we now express the noncommutativity of the algebra in the following way  
\begin{eqnarray}
&& 
P^{R\Lambda\tau\alpha}
P^{R^{\prime}\Lambda^{\prime}\tau^{\prime}\alpha^{\prime}}
\nonumber \\
&=&
\left(
\sum_{i}
D^R_{ab}(\sigma^i)
\tilde{S}^{RR,\Lambda\tau,\alpha}_{ba}\sigma_i
\right)
\left(
\sum_{k}
D^{R^{\prime}}_{cd}(\sigma^k)
\tilde{S}^{R^{\prime}R^{\prime},\Lambda^{\prime}\tau^{\prime},\alpha^{\prime}}_{dc}
\sigma_k
\right)
\nonumber \\
&=&
\frac{1}{d_R}\delta_{RR^{\prime}}
\sum_{i}
D^{R}_{cb}(\sigma^{i})
(\tilde{S}^{RR,\Lambda\tau,\alpha }
\ast
\tilde{S}^{R^{\prime}R^{\prime},\Lambda^{\prime}\tau^{\prime},\alpha^{\prime}})_{bc}
\sigma_i,
\label{covariant_noncommutative_product}
\end{eqnarray}
where we have defined 
\begin{eqnarray}
(\tilde{S}^{RR,\Lambda\tau,\alpha }
\ast
\tilde{S}^{R^{\prime}R^{\prime},\Lambda^{\prime}\tau^{\prime},\alpha^{\prime}})_{bc}
=
\tilde{S}^{RR,\Lambda\tau,\alpha }_{ba}
\tilde{S}^{R^{\prime}R^{\prime},\Lambda^{\prime}\tau^{\prime},\alpha^{\prime} }_{ac}.
\end{eqnarray}
This non-commutative product is an analogue of  
$P^R_{A,\mu\nu}P^S_{B,\alpha\beta}=
\delta_{\nu\alpha}\delta_{RS}\delta_{AB}P^S_{B,\mu\beta}$.

Defining 
\begin{eqnarray}
\hat{D}_{ab}^{R}
:=d^R
D^{R}_{ba}(\sigma^i)\sigma_i, 
\end{eqnarray}
the product rule is the usual matrix product\footnote{
We have used the formula
\begin{eqnarray}
\sum_{i}D^{R^{\prime}}(\sigma^i)_{cb}\sigma_i
\sum_{j}
D^{R}(\sigma^j)_{ef}\sigma_j
=
\frac{1}{d_R}\delta^{RR^{\prime}}
\sum_{i}D^{R}(\sigma^i)_{eb}
\sigma_i\delta_{cf}.
\end{eqnarray}
}
\begin{eqnarray}
\hat{D}^{R}_{ab}\hat{D}^{R^{\prime}}_{cd}
=\delta_{bc}\delta^{RR^{\prime}}\hat{D}^{R}_{ad}.
\end{eqnarray}
With this notation we have
\begin{align}
P^{R\Lambda\tau\alpha}
=\frac{1}{d_R}\hat{D}^R_{ab}
\tilde{S}^{RR,\Lambda\tau,\alpha}_{ab},
\end{align}
\begin{eqnarray}
P^{RR,\Lambda\tau,\alpha}
P^{R^{\prime}R^{\prime},\Lambda^{\prime}\tau^{\prime},\alpha^{\prime}}
=
\frac{1}{(d_R)^2}
\delta_{RR^{\prime}}
\hat{D}^{R}_{ab}
(\tilde{S}^{RR,\Lambda\tau,\alpha }
\ast
\tilde{S}^{R^{\prime}R^{\prime},\Lambda^{\prime}\tau^{\prime},\alpha^{\prime}})_{ab},
\end{eqnarray}
\begin{align}
&
P^{R\Lambda\tau\alpha}
P^{R^{\prime}\Lambda^{\prime}\tau^{\prime}\alpha^{\prime}}
P^{R^{\prime\prime}\Lambda^{\prime\prime}\tau^{\prime\prime}\alpha^{\prime\prime}}
\nonumber \\
&=
\frac{1}{(d_R)^3}
\delta_{RR^{\prime}}\delta_{RR^{\prime\prime}}
\hat{D}^{R}_{ab}
(
\tilde{S}^{RR,\Lambda\tau,\alpha}
\ast
\tilde{S}^{RR,\Lambda^{\prime}\tau^{\prime},\alpha^{\prime}}
\ast
\tilde{S}^{RR,\Lambda^{\prime\prime}\tau^{\prime\prime},\alpha^{\prime\prime}}
)_{ab}.
\end{align}
The closed string state can be calculated to be
\begin{eqnarray}
\Pi([\sigma_s]_H)&=&
\sum_{R,\Lambda,\tau,\alpha}
\sum_{\Lambda^{\prime},\tau^{\prime},\alpha^{\prime}}
\sum_{\Lambda^{\prime\prime},\tau^{\prime\prime},\alpha^{\prime\prime}}
D^R_{cd}(\sigma_s)
\tilde{S}^{RR,\Lambda^{\prime}\tau^{\prime},\alpha^{\prime}}_{cd}
(
\tilde{S}^{RR,\Lambda\tau,\alpha}\ast
\tilde{S}^{RR,\Lambda^{\prime\prime}\tau^{\prime\prime},\alpha^{\prime\prime}}
)_{ff}
\nonumber \\
&&\frac{1}{d_R}
\hat{D}^{R}_{ab}
(
\tilde{S}^{RR,\Lambda\tau,\alpha}\ast
\tilde{S}^{RR,\Lambda^{\prime}\tau^{\prime},\alpha^{\prime}}\ast
\tilde{S}^{RR,\Lambda^{\prime\prime}\tau^{\prime\prime},\alpha^{\prime\prime}})_{ab}.
\label{pi_sigma_s_evaluated}
\end{eqnarray}

\if
We can also show that 
\begin{eqnarray}
(
\tilde{S}^{R,A}\ast\tilde{S}^{R,A^{\prime}}\ast\tilde{S}^{R,A^{\prime\prime}})_{mb}
g^R_{AA^{\prime\prime}}
=\frac{1}{|H|}\sum_h D^{R}(h)_{mb}D^{R}(h^{-1})_{cd}\tilde{S}^{R,A^{\prime}}_{dc}
\label{SSSg=DDS}
\end{eqnarray}
which comes from 
\begin{eqnarray}
\sum_A\tilde{S}^{R,A}_{ac}\tilde{S}^{R,A}_{db}
=\frac{1}{|H|}\sum_h D^{R}(h)_{ab}D^{R}(h)_{cd}
\end{eqnarray}

\begin{eqnarray}
\Pi( P^{R\Lambda_1\tau_1\alpha_1})&=&
\sum_{\Lambda,\tau,\alpha}
\sum_{\Lambda^{\prime\prime},\tau^{\prime\prime},\alpha^{\prime\prime}}
(
\tilde{S}^{RR,\Lambda\tau,\alpha}\ast
\tilde{S}^{RR,\Lambda^{\prime\prime}\tau^{\prime\prime},\alpha^{\prime\prime}}
)_{ff}
\nonumber \\
&&\frac{1}{(d_R)^2}
\hat{D}^{R}_{bd}
(
\tilde{S}^{RR,\Lambda\tau,\alpha}\ast
\tilde{S}^{RR,\Lambda_1\tau_1,\alpha_1}\ast
\tilde{S}^{RR,\Lambda^{\prime\prime}\tau^{\prime\prime},\alpha^{\prime\prime}})_{bd}
\label{closed_string_state_covariant}
\end{eqnarray}
\note{(If everything is consistent, the RHS is a basis of the centre.)}

Using (\ref{SSSg=DDS}), 
\begin{eqnarray}
tr^{(r)}(\Pi( \bar{f}^{iR\Lambda_1\tau_1\alpha_1}
\sigma_i ))
&=&\frac{1}{d_R}\frac{1}{|H|}\sum_h \chi^{R}(h)D^{R}(h^{-1})_{cd}
\tilde{S}^{RR,\Lambda_1\tau_1,\alpha_1}_{dc}
\nonumber \\
&=&\frac{1}{d_R}\frac{1}{|H|}\sum_h \sum_A M^R_A \chi_A(h)
D^{R}(h^{-1})_{cd}\tilde{S}^{RR,\Lambda_1\tau_1,\alpha_1}_{dc}
\nonumber \\
&=&\frac{1}{d_R} \sum_A \frac{1}{d_A} M^R_A D^{R}(p_A)_{cd}
\tilde{S}^{RR,\Lambda_1\tau_1,\alpha_1}_{dc}
\nonumber \\
&=&\frac{1}{d_R}\sum_A \frac{1}{d_A} M^R_A D^{R}(p_A)_{cd}
\tilde{S}^{RR,\Lambda_1\tau_1,\alpha_1}_{dc}
\end{eqnarray}
we have used $\chi^{R}(h)=\sum_A \chi^R_A(h)=\sum_A M^R_A \chi_A(h)$. 
Note that $D^{R}(p_A)_{cd}$ is zero if $c\neq d$.\footnote
{
\begin{eqnarray}
&&
D^{R}(p_A)_{cd}
\nonumber \\
&=&
\langle R,c|p_A |R,d \rangle
\nonumber \\
&=&
\sum_{m,\mu}\sum_{m^{\prime},\nu}
\langle R,c|R\rightarrow A,m,\mu \rangle 
\langle R\rightarrow A,m,\mu |
p_A |R\rightarrow A,m^{\prime},\nu \rangle
\langle R\rightarrow A,m^{\prime},\nu|
R,d \rangle
\nonumber \\
&=&
\sum_{m,\mu}
\langle R,c|R\rightarrow A,m,\mu \rangle 
\langle R\rightarrow A,m,\mu|
R,d \rangle
\end{eqnarray}
}                                         
So only the diagonal part $c=d$ in $\tilde{S}^{RR,\Lambda_1\tau_1,\alpha_1}_{dc}$ 
contributes to the closed string variable. 

It is also helpful to see the basis change between 
the restricted Schur and the covariant given in [(2.5), 0810.4217]. 
From [(2.5), 0810.4217], we find  
\begin{eqnarray}
P^R_{A}\propto (P^R_A)_{ba}\tilde{S}^{RR,\Lambda\tau,\alpha}_{ab}
\bar{f}^{iR\Lambda\tau\alpha}
\sigma_i
\end{eqnarray}
(\ref{closed_string_state_covariant}) is non-zero for 
$R,\Lambda,\tau,\alpha$ that give $a=b$. 
(Exactly speaking, 
this statement would be not correct..) 
\fi


\end{document}